# Software quality: A Historical and Synthetic Content Analysis

Peter Kokol
University of Maribor, Faculty of Electrical Engineering and Computer Science, Maribor, Slovenia

**Abstract**
**Background:** Interconnected computers and software systems have become an indispensable part of people's lives, therefore software quality research is becoming more and more important. There have been multiple attempts to synthesize knowledge gained in software quality research, however, they were focused mainly on single aspects of software quality and not to structure the knowledge in a holistic way. The aim of our study was to close this gap. **Methodology:** The software quality publications were harvested from the Scopus bibliographic database. The metadata was exported first to CRexlporer, which was employed to identify historical roots, and next to VOSViewer, which was used as a part of the synthetic content analysis. In our study we defined synthetic context analysis as a triangulation of bibliometrics and content analysis. **Results:** Our search resulted in 14451 publications. The performance bibliometric study showed that the production of research publications relating to software quality is currently following an exponential growth trend and that the software quality research community is growing. The most productive country was the United States and the most productive Institution The Florida Atlantic University. The synthetic content analysis revealed that the published knowledge can be structured into 10 themes, the most important being the themes regarding software quality improvement with enhancing software engineering, advanced software testing, and improved defect and fault prediction with machine learning and data mining. According to the analysis of the hot topics, it seems that future research will be directed into developing and using a full specter of new artificial intelligence tools (not just machine learning and data mining) and focusing on how to assure software quality in agile development paradigms. **Conclusion:** The study can inform software developers, theoreticians, practitioners, software users, and all other interested stakeholders about the main aspects and themes of software quality research, enable them to gain new insight into the topic or deepen their knowledge in specific aspects of software quality.

Short title: Software quality

**Conflict statement:** Authors declare no conflicts of interest:
**Funding:** The study presented in this paper was partially  funded by the Slovenian Research agency by the grant J7-2605

## Introduction

Interconected computers and software systems have become an indispensable part of peoples lives and the necessary toll for performing their daily personal and business obligations and activities. To fulfill global user needs for storing, retrieving and processing information, knowledge and wisdom those systems have to be supported by quality software, which should function correctly and reliably, be easy, safe and fit to use, test, reuse and maintain, and finally to conform to stakeholders requirements. Therefore, software quality is not only one of the most important, but also multidimensional attributes of computer software (1–4).

Consequently there have been multiple attempts to synthetise knowledge gained in software quality research, however they were focused mainly on single aspects of software quality, like measurement (5), human-software interaction (6), empirical analysis of Code smells and refactoring(7), design patterns (8). Quality models (9), human factors (10), testing (11) and quality prediction (12,13) or to specific development approaches like agile (14). Bibliometrics (15) is another approach to synthetises knowledge (16). Similarly to above the only two bibliometrics studies concerned with software quality were focused on just specific aspects of software quality, one on defect prediction (17) and the other on code smells (18).

To close the gap regarding the lack of holistic knowledge synthesis studies of software quality research we performed a historical and performance bibliometric analysis and synthetic bibliometric mapping-based content analysis. In that manner, our aim was to identify the most productive countries and institutions, most prolific source titles, publication production trends, historical roots and hot topics, and structure the research content into themes.

The choice to focus on the field of software quality was motivated by the belief that software quality assurance is an extremely important factor in software development for a myriad of stakeholders like software developers, theoreticians, practitioners and of course software users. The study can help them to gain new insights into the topic, deepen their knowledge or inform them about the trends and important themes in software quality research. To the best of our knowledge, a similar bibliometric study that provides an overview of the current state and development of software quality research from its beginnings to the present has not been conducted so far. Therefore, through the present study, we eliminate this shortcoming and fill in the current gap.

## Methodology

The search was performed in Scopus (Elsevier, Netherlands), the largest abstract and citation database of peer-reviewed literature. The corpus was formed on March 5th, 2021, for the whole period covered by Scopus. We used the search string *"quality software" OR "quality of software" OR "software quality"* in information source titles, abstracts, and keywords. The reliability of the search was calculated using the precision (fraction of the documents retrieved that are relevant to the user's information need) and recall (fraction of the documents that are relevant to the query that are successfully retrieved) information retrieval functions using 20 important software quality publications and 10 eminent authors. All authors and publications were retrieved using the above search string. Using Scopus built in functions we extracted the publication titles, authors` affiliations` details, source title, publication type, abstracts, publishing years, author keywords, number of citations, and references and stored them in the CSV formatted corpus file.

### *Historical roots*

Historical roots are important publications in a specific research area (19). To identify them we used a specialized software, namely CitedReferenceExplorer or CRE in

short(www.crexplorer.net). CRE on the so called Reference Publication Year Spectroscopy (RPYS) approach (20). It has been successfully used in different research areas (21–24). We used the CRE tabular output to identify historical roots for the period till the year 1955, the CRE graphical output (spectrogram) after that and the N_TOP10 indicator (29), to identify publication which were 10% of most cited publication over at least 20 years. To perform the analysis we imported the corpus metadata to the CRE software. CRE default values were used for the analysis.

***Synthetic content analysis***

Another approach to deal with the explosive growth in the research literature production is knowledge synthesis. Knowledge synthesis roots dates back more than 120 years, however become more popular in the 1960s (30), and even more commonly used toward the end of the millennia with the emergence of the evidence-based paradigm. (31,32). One of the more popular knowledge synthesis methods usable in both qualitative and quantitative research is content analysis. Its main advantages are that it is content-sensitive, highly flexible, and can be used to analyse many types of data, either in inductive or deductive manner (33).

To enable the knowledge synthesis of a ten thousands of publications Kokol et al (34) triangulating bibliometrics, text mining and content analysis into synthetic content analysis which was used in our study to structure and map the software quality research literature as presented bellow:

1. Harvest the research publications concerning software quality to represent the content to analyse.
2. Condense and code the content using text mining of publication abstracts and author keywords. Authors keywords were selected as candidate codes, cause they most concisely present the content of a publication (35)

3. Map the codes using bibliometric mapping and induce a clustered author keywords landscape.
4. Analyse the connections between the codes in individual clusters and map them into categories.
5. Analyse sub-categories and label each cluster with an appropriate theme.

**Results and discussion**

The search resulted in 14451 publications. Among them, there were 8891 conference papers, 4498 articles, 390 conference reviews, 256 reviews, 68 editorials, 57 books 29 short papers and 262 other types of publications.

***Performance and productivity bibliometrics***

The most productive countries were United States (n=2815), China (n=1567), India (n=1297), Germany (n=907 and Canada (n=681) publications. The most productive instructions were Florida Atlantic University (n=225), Beihang University (n=107), Peking University (n=93), Amity University (n=88), and San Paulo University (n=82). The most prolific source titles are Lecture Notes in Computer Science (n=714), ACM International Proceedings Series (n=349), Proceedings International Conference on Software Engineering (n=313), Communication in Computer and Information Science (n=211) and CEUR Workshop Proceedings (n=206). The most prolific journal is Information and Software Technology on sixth place with 197 publications, followed by the Software Quality Journal on seventh place with 184 publications. The first paper indexed in Scopus was published in 1954 and the second in 1970 (Figure 1.). After then the production was low, not exceeding five publications per year. The linear growth trend begun in 1978 and the exponential growth in 2001, reaching its peak in 2019 with 1027 publications. The trend of the number of proceedings and journal papers followed the similar pattern till 2001, then the trend split. The number of journal papers followed a steady exponential trend reaching 400 articles in 2020, while the number of conference papers had an explosive growth period from 2001 till 2007, after that the number of papers reminded steady till 2019 when it reached its peak with 701 papers. According to the above trends we can identify three milestones in software quality research, namely (1) beginning of seventies when first software crisis emerged (36), (2) end of the seventies when software quality models and measurements started to gain in importance (37), and finally (3) in the beginning of the new century with the advent of agile programing (38).

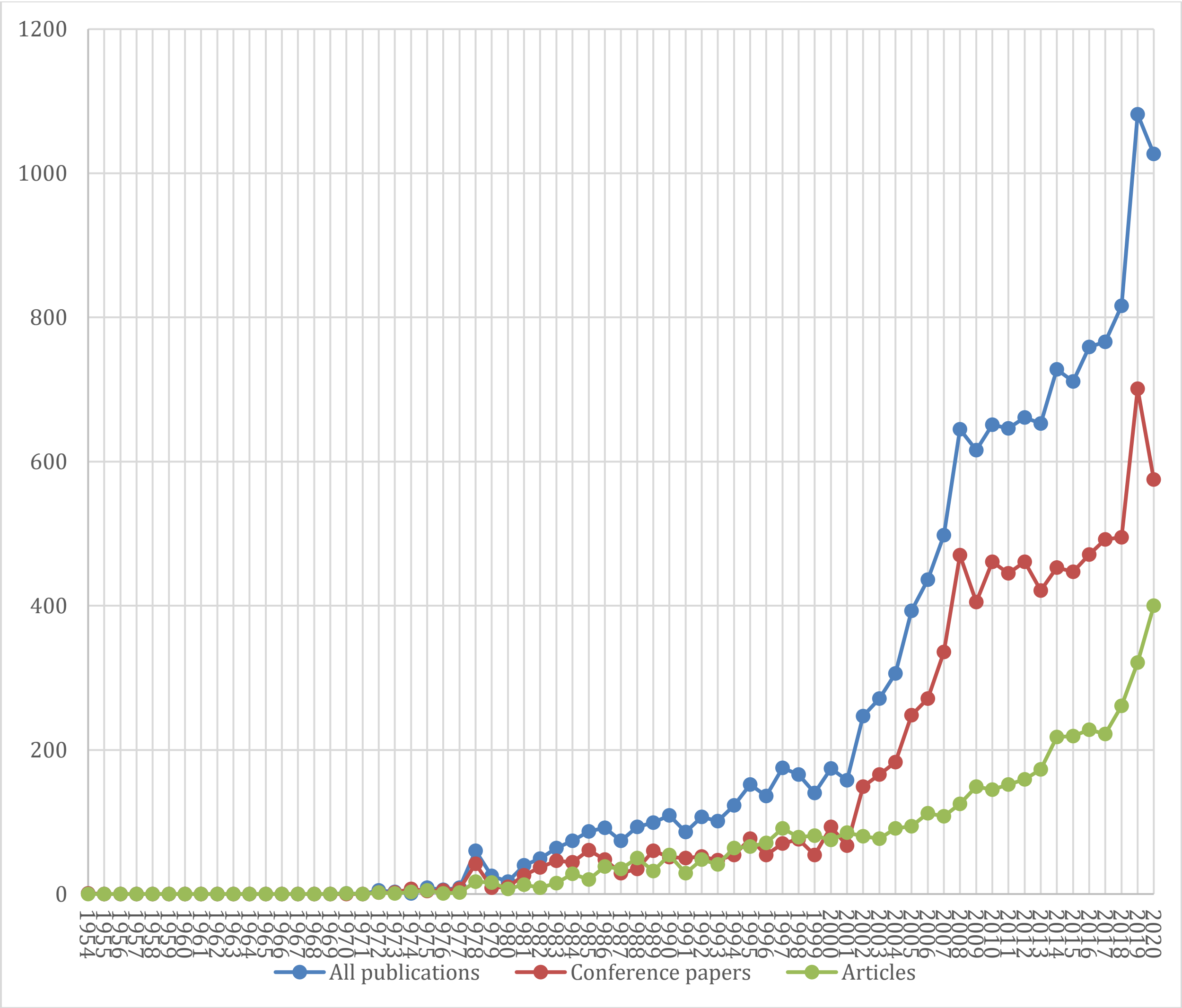

Figure 1. The dynamics of the software quality research expressed by the number of peer reviewed publications

***Historical roots***

The results of the CRE analysis are shown in Table 1. The oldest historical root dates back to the year 1735. The historical roots in the “pre-software era” (1763-1971) were mainly concerned with statistical analysis. The exceptions were Dewesy book on teaching scientific logic to pupils (39), the Fleshs paper presenting the readability formula of literary texts (40) and the Millers paper on capacity limits of processing information (41). We may observe that the early historical roots represent the scientific core of software quality, and origins of code readability and modularity. The next period historical roots (1972-1992) were concerned with the quality of software developed in the scope of the procedural languages paradigm, and the last period historical roots with the quality of object-oriented software.

Table 1. Historical roots

| Publishing year | Title | Authors |
|---|---|---|
| 1735 | Systema naturae | Linnaeus, C. |
| 1763 | An essay towards solving a problem in the doctrine of chances Philosophical Trans. Royal Soc. of London, 53 | Bayes, T. |
| 1880 | On the diagrammatic and mechanical representation of propositions and reasonings | Venn, J. |
| 1901 | On lines and planes of closest fit to systems of points in space | Pearson, K. |
| 1904 | The proof and measurement of association between two things | Spearman, C. |
| 1910 | How We Think | Dewey, J. |
| 1921 | On the probable error of a coefficient of correlation deduced from a small sample | Fisher, R.A. |
| 1932 | A technique for the measurement of attitudes | Likert, R. |
| 1938 | A new measure of rank correlation | Kendall, M.G. |
| 1945 | Individual comparisons by ranking methods | Wilcoxon, F. |
| 1947 | On a test of whether one of two random variables is stochastically larger than the other | Mann, H.B.; Whitney, D.R. |
| 1948 | A new readability yardstick | Flesch, R.F. |
| 1956 | The magical number seven, plus or minus two: Some limits on our capacity for processing information | Miller, G. |
| **1959** | Convergent and discriminant validation by the multi-trait-multimethod matrix | Campbell, D.; Fiske, D. |
| **1972** | On the criteria to be used in decomposing systems into modules | Parnas, D.L. |
| **1976** | A complexity measure | McCabe, T. |
| **1976** | Design and code inspections to reduce errors in program development | Fagan, M.E. |
| **1977** | Elements of Software Science | Halstead, M.H. |
| **1977** | Factors in Software Quality | McCall, J.; Richards, P.; Walters, G. |
| **1978** | Characteristics of Software Quality | Boehm, B.W.; Brown, J.R.; Kaspar, H.; Lipow, M.; Macleod, G.J.; Merrit, M.J. |
| **1981** | Software Engineering Economics | Boehm, B.W. |
| **1993** | Object-oriented metrics that predict maintainability | Li, W.; Henry, S. |
| **1994** | A metrics suite for object oriented design | Chidamber, S.R.; Kemerer, C.F. |
| **1996** | A validation of object-oriented design metrics as quality indicators | Basili, V.R.; Briand, L.C.; Melo, W.L. |
| **2002** | A hierarchical model for object-oriented design quality assessment | Bansiya, J.; Davis, C.G. |
| **2003** | Empirical analysis of ck metrics for object-oriented design complexity: Implications for software defects | Subramanyam, R.; Krishnan, M. |
| **2005** | Empirical validation of object-oriented metrics on open source software for fault prediction | Gyimothy, T.; Ferenc, R.; Siket, I. |
| **2007** | Data mining static code attributes to learn defect predictors | Menzies, T.; Greenwald, J.; Frank, A. |

### *Synthetic content analysis*

The author keywords landscape is shown in Figure 2 and the results of the synthetic thematic analysis in Table 2. Ten clusters and consequently 10 themes emerged. Those themes were derived from 93 most frequent codes and 46 categories. The most important themes regarding the most frequent codes related to the theme are *Improving software quality with advanced software engineering* (42–44), *Software testing* (45–47) and *Using software metrics as input to machine learning and data mining to predict and classify faults and defects* (48–50).

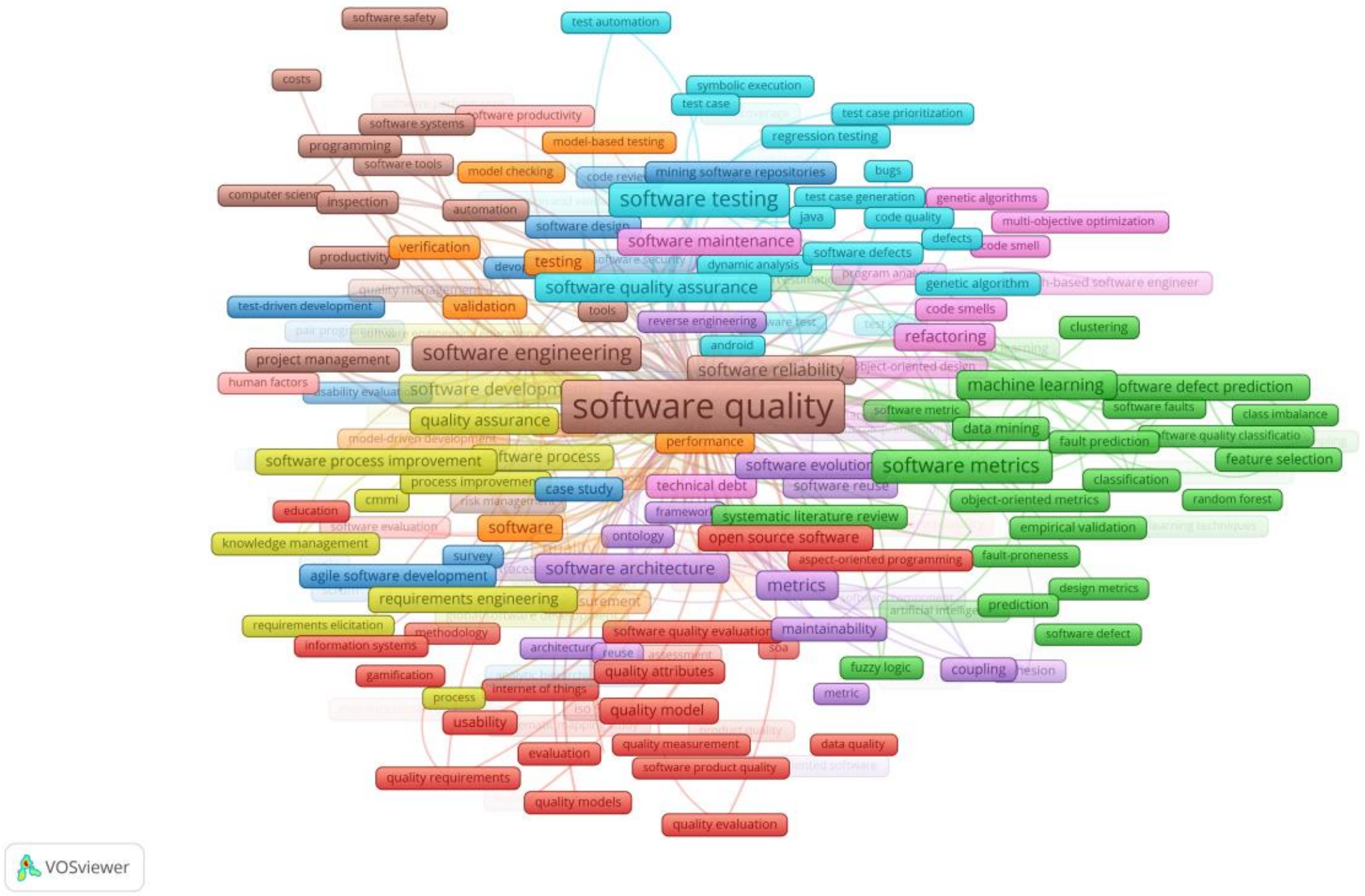

Figure 2. Authors keywords cluster landscape

Table 1: Software quality research themes (numbers in parenthesis present the number of publications in which an author keyword appeared)

| *Colour* | *More frequent codes* | *Categories* | *Themes* |
|---|---|---|---|
| *Yellow* | Software process improvement (137), Requirements engineering (113), quality assurance (152), software development (210), CMM (70), Knowledge management (8) | Software quality management employing software process improvement with CMM, Improving requirements engineering, Quality assurance in software development, Using knowledge management in software development | Software quality assurance in a scope of a mature software development process |
| *Rose* | Software evaluation (28), Human factors (30), Software complexity (25), Software productivity (22) | Human factors in software evaluation | Human factors in software evaluation |
| *Brown* | Software quality (2491), Software engineering (579), Software reliability (209), Project management | Improving software quality and productivity with software engineering, Reliability | Improving software quality with advanced software engineering |
| | (84), Software measurement (90), Quality management (55), Programming (51), Productivity (50), Inspection (38), Automation (36) | measurement to increase software safety, Integration software quality in project management to reduce costs, Computer science of software measurement. Improving software quality with inspections, Tools for automation of software engineering | |
| *Dark blue* | Empirical study (141), Agile software development (76), Empirical software engineering (75), Mining software repositories (50), Software design (62), Extreme programming (38) Software Security (35), Code revive (35), DevOps (32) | Empirical software engineering in agile based software design paradigm, Mining software repositories in empirical analysis of code reviews, Test driven development and extreme programming, | Empirical analysis of agile approaches in software development |
| *Red* | Quality model (117), Open source software (112), Quality attributes (69), Usability (99), Cloud computing (59), Non-functional requirements (56), Quality metrics (53), Quality evaluation (51), Software maintainability (48), ISO9126 (40), Quality assessment (38), ISO/IEC 25010 (33), Static code analysis (30), Data quality (29), Software product quality (24) | Quality evaluation and assessment based on quality models, quality attributes, metrics and standards, Quality evaluation of open source software, Improving software maintainability with aspect oriented programming, Quality evaluation of usability and user experience, Quality of nonfictional requirements, Quality of service-oriented architectures/cloud computing | Total quality evaluation of connected software systems, based on general quality models and standards |
| *Orange* | Quality (214), Software (178), Testing (138), Reliability (106), Measurement (79), UML (76), validation (70), Verification (76), Security (57), Formal methods (51), Performance (49), Embedded software (42) | Measuring, testing and validation of software quality/reliability/security, Software design with the Model-driven engineering approach, Safety and reliability of software, Reliability of embedded software, UML in model driven development, Testing, validation and verification of software, | Managing the quality of embedded software |
| | | Formal methods in software engineering, Performance and reliability measurement of sofware | |
| *Light blue* | Software testing (543), Software quality assurance (260), Static analysis (100), Regressions testing (63), Genetic algorithm (63), Software defects (44), Unit testing (39), Test case prioritisation (35), Bugs (34), Automated testing (33), Mutation testing (32), Verification and validation (30), Code quality (30), Test case generation (29), Dynamic analysis (24); | Software quality assurance with testing, Software quality assurance using verification, validation and static analysis, Static and dynamic analysis, to assure code quality, Test case prioritisation in regression testing, Use o genetic algorithms in regression and mutation testing; Improving quality of unit tests with mutation testing, Software test automation, | Software testing |
| *Pink* | Software maintenance (177), Refactoring (162), Technical debt (93), Code smells (66), Search based software engineering (42), Multi-objective optimization (28) | Improving software maintenance by removing smelled code with refactoring, Dealing with technical debt with code smells identification, Multi-objective optimisation of software with genetic algorithms in search based software engineering | Improving software maintenance with refactoring and search based software engineering |
| *Viollet* | Software architecture (261), Metrics (245), Software evolution (122), Software reuse (89), Design pattern (83), Coupling (68), Ontology (62), Cohesion (45), Component (45), Traceability (41), Software product line (39) | Software evolution based on software architectures and design patterns, using coupling and cohesion metrics to improve maintainability, Software evolution with reverse engineering, Reducing complexity with coupling and cohesion metrics, Using components for software reuse, Use of ontologies to trace software architecture decisions, Implementing reuse using software architecture and software product lines | Software architectures and software evolution |
| *Green* | Software metrics (493), Machine | Using software metrics as input to | Using software metrics as input to |

| | | |
|---|---|---|
| learning (234), Software defect prediction (143), Data mining (131), Software fault prediction (104), Systematic literature review (74), Fuzzy logic (66), Clustering (49), Deep learning (43) | machine learning to predict faults and defects, Using machine learning and data mining for fault prediction and classification, Knowledge synthesis on software quality prediction with systematic literature reviews | machine learning and data mining to predict and classify faults and defects |

### ***Hot topics***

To identify hot topics we extended the methodology developed by Kokol et al (51) with the synthetic content analysis and used this extended approach to compare the corpus of publications published in 2018 and 2019 with the corpus of publications published in 2020 and 2021. All the new categories and themes emerging in 2020/21, and the categories and themes emerging in 2018/19 mostly cited in 2020/21 were recognised as hot topics. In that manner we identified:

- A new theme: Improving software development with the Integration of CMMI into agile approaches (52).
- New categories:
    - Naturale language processing of software documents to elicit high quality requirements (53).
    - Software quality attributes in agile environments (54).
    - Software architectures for internet of things (55).
    - Software maintenance of blockchain based software (56).
- Most cited categories in 2020/21:
    - Detecting code smells with genetic algorithms (57).

- Use of Artificial intelligence in risk management (58).
- Use of Multi criteria decision making in software quality modeling (59).

***Strengths and limitations***

The main strength of the study is that it is the first holistic synthetic content analysis of the small dataset research. One possible limitation is that the analysis was limited to publications indexed in Scopus only, however due to the fact that Scopus covers the largest and most complete set of information titles, it is our belief that we analysed most of the important peer reviewed publications.

## Conclusions

Our bibliometric study showed that production of research publications relating to software quality is currently following an exponential growth trend and that software quality research community is growing. The synthetic content analysis revealed that the published knowledge can be structured into 10 themes most important being the themes regarding software quality improvement with enhancing software engineering, advanced software testing and improved defect and fault prediction with machine learning and data mining. According to the hot topics analysis it seems that future research will be directed into developing and using a full specter of new artificial intelligence tools (not just machine learning and data mining) and focusing on how to assure software quality in agile development paradigms.